\documentclass[12pt,onecolumn]{IEEEtran}
\usepackage[colorlinks,bookmarksopen,bookmarksnumbered,citecolor=red,urlcolor=red,]{hyperref}
\usepackage{indentfirst}
\usepackage{algorithm,algorithmic,amsbsy,amsmath,amssymb,epsfig,bbm,mathrsfs, bbm} 
\usepackage{multirow}
\usepackage{mathrsfs}
 \usepackage{epstopdf}

\usepackage{verbatim}
\usepackage{amsfonts}
\usepackage{amsthm}
\begin{document}

\title{Economics of Internet of Things (IoT): An Information Market Approach}
\author{\vspace{0.2in}
{\em Open Call}	
\vspace{0.2in} 
\\ Dusit Niyato, School of Computer Engineering, Nanyang Technological University (NTU), Singapore	\\
Xiao Lu, Department of Electrical and Computer Engineering, University of Alberta, Canada	\\
Ping Wang, School of Computer Engineering, Nanyang Technological University (NTU), Singapore	\\
Dong In Kim, School of Information and Communication Engineering, Sungkyunkwan University (SKKU), Korea \\
Zhu Han, Electrical and Computer Engineering, University of Houston, Texas, USA.
\thanks{ Contact: \textbf{D. I. Kim}, School of Information and Communication Engineering, Sungkyunkwan University (SKKU), 23528 Engineering Building \#1, Suwon, Korea 440-746, Tel: +82-31-299-4585, Fax: +82-31-299-4673 E-mail: dikim@skku.ac.kr.  Editor: Abderrahim Benslimane } 
}

\maketitle

\begin{abstract}
Internet of things (IoT) has been proposed to be a new paradigm of connecting devices and providing services to various applications, e.g., transportation, energy, smart city, and healthcare. In this paper, we focus on an important issue, i.e., economics of IoT, that can have a great impact to the success of IoT applications. In particular, we adopt and present the information economics approach with its applications in IoT. We first review existing economic models developed for IoT services. Then, we outline two important topics of information economics which are pertinent to IoT, i.e., the value of information and information good pricing. Finally, we propose a game theoretic model to study the price competition of IoT sensing services. Some outlooks on future research directions of applying information economics to IoT are discussed.
\end{abstract}

\begin{IEEEkeywords}
Internet of Things (IoT), economic model, information economics, game theory
\end{IEEEkeywords}

\section{Introduction}

Internet of things (IoT) is a new paradigm of connecting objects through the Internet. Devices and people will have ability to transfer data over wired and wireless networks with minimal human intervention. Devices can be sensors and actuators that generate data and receive instruction to perform certain sets of functions. Thus, IoT has a great potential to facilitate domain-specific usage and to improve performances of the systems in many applications such as transportation, energy management, manufacturing, and healthcare~\cite{atzori2010}. IoT integrates several technologies, e.g., hardware design, data communication, data storage and mining, information retrieval and presentation. It also involves many disciplines including engineering, computer science, business, social science, etc., to achieve goals of target applications. Therefore, designing and developing IoT systems and services require holistic approaches including engineering and management that ensure efficiency and optimality in every part of IoT.

In this paper, we focus particularly on the economics aspect of IoT. Economic issues include cost-benefit analysis, user utility, and pricing. We first highlight the factors that make economic issues imperative for IoT, and then review related works of economic models developed for IoT services and applications. Next, we discuss a potential approach, i.e., information economics, and its applications in IoT. Specifically, two major directions are presented, i.e., the value of information and information good pricing. Finally, we present a demonstrative economic model based on game theory to study IoT sensing service competition. We show the effects of substitute (and complementary) services on the equilibrium prices that users can use one (and all of services) to obtain sensing information, respectively. Finally, open research directions are outlined.

The remainder of this article is organized as follows. Section~\ref{sec:economic_models} presents a general structure of IoT and discusses the economic issues. Section~\ref{sec:infoecon} introduces the concept of information economics and its potential applications in IoT. Then, Section~\ref{sec:competition} proposes a game theoretic model to analyze price competition of IoT sensing services. Finally, Section~\ref{sec:conclusion} concludes the article.

\section{Economic Models of Internet of Things}
\label{sec:economic_models}

This section first introduces an overview of IoT. Then, we discuss some economic issues and techniques used in IoT.

\subsection{Internet of Things}

\begin{figure}
\begin{center}
$\begin{array}{c} \epsfxsize=3.0 in \epsffile{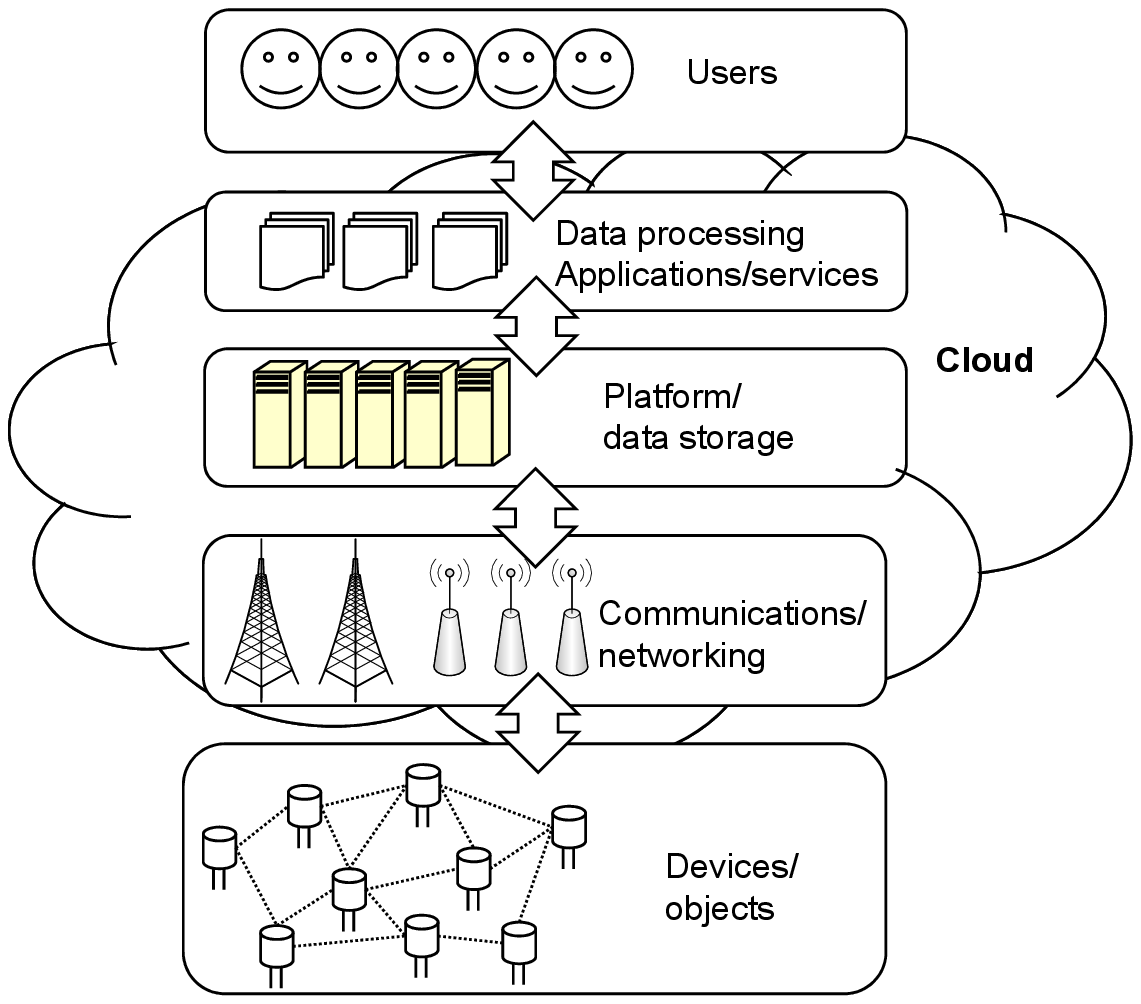} \\ [-0.2cm]
\end{array}$
\caption{Internet of Things (IoT) representative model.} 
\label{fig:iot_model}
\end{center}
\end{figure}

\vspace{0.2in}
\noindent{\bf Callout: Figure~\ref{fig:iot_model} Internet of Things (IoT) representative model.}
\vspace{0.2in}

IoT is a board concept introduced to describe a network of things or objects. The objects can be sensors, actuators, electronic devices, etc., that are able to connect to the Internet through wireless and wired connections. Figure~\ref{fig:iot_model} shows the representative structure of IoT~\cite{gluhak2011}. IoT can be divided into different tiers so that the system is scalable and able to support heterogeneous environment with high flexibility and reliability. 

\begin{itemize}
	\item {\em Devices:} This perception and action layer is composed of low-level devices such as sensors and actuators. They have limited computing, data storage, and transmission capability. Thus, they perform only primitive tasks such as monitoring environment conditions, collecting information, and changing system parameters. Basically, the devices are the end-point of information, i.e., sources or sinks, in IoT. They are generally connected with Internet gateways for data aggregation. They can also be connected among each other with peer-to-peer connections for information forwarding. 

\item {\em Communications and Networking:} This layer provides data communications and networking infrastructure to transfer data of devices efficiently. Typically, wireless networks are used to connect the devices, which can be mobile or fixed, to the gateways. The data is transferred from gateways to the Internet via backbone networks such as mesh networks. 

\item {\em Platform and Data Storage:} This layer provides facility for data access and storage. It can be hardware and platform in local data centers or services in the cloud, e.g., Infrastructure-as-a-Service (IaaS) and Platform-as-a-Service (PaaS).

\item {\em Data Management and Processing:} This software layer provides services for users to access functions of IoT services. It is composed of backend data processing, e.g., database and decision unit, and frontend user and Business-to-Business (B2B) interfaces.

\end{itemize}

The resource management will be an important issue for delivering efficient IoT services to users. Different resources have to be optimized to minimize the cost, to maximize the utilization and profit, as well as to satisfy Quality of Service (QoS) of IoT services~\cite{xu2014}. Different layers involve different resources, e.g., energy used for the devices to operate, spectrum and bandwidth for wireless and wired networks to transfer data, computing and data storage for the platform and infrastructure, and data processing services for IoT applications. 

For example, in the IoT-based home surveillance applications, video cameras and motion sensors operated on a battery are deployed at different locations in a house. The cameras and sensors transfer data back to the gateway via wireless connections. The video and sensing data are stored in the cloud and is processed to detect whether there is an intrusion or not. If there is an intrusion, the service will stream video data to the end user's devices and inform security officers for further action. In this example, for the cameras and sensors, energy from the battery and wireless transmission bandwidth are scarce resources to be optimized to meet delay and reliability requirements. Cloud data storage and computation services, e.g., a virtual machine hosting, have to be allocated for signal detection and image processing. Mobile services to stream video traffic can be regarded as a resource that needs to be acquired.

Typical approaches of solving resource allocation problems in IoT are based on system optimization, e.g.,~\cite{xu2014}. In the system optimization-based resource allocation, the system has one objective with constraints. The system is able to control the resource usage to achieve the optimal solution that maximizes/minimizes the objective while meeting all the constraints. For example, in~\cite{xu2014}, the system optimization for time slot allocation to support multi-camera video streaming under IoT services is proposed. The objective is to maximize the sensing utility by adjusting the data transmission rate, which is the function of time slot. The constraints are to ensure the delay deadline of video traffic. Its optimal solution is obtained based on convex optimization.

\subsection{Economic Issues and Incentive Approaches}

Traditional system optimization may not be suitable for IoT in many circumstances because of the following reasons.
\begin{itemize}
	\item {\em Heterogeneous Large-Scale Systems:} As shown in Fig.~\ref{fig:iot_model}, IoT usually involves and consists of a number of diverse components, e.g., several thousands of sensors, hundreds of access points, and tens of cloud data centers, integrated in a highly complex manner. Thus, the centralized management approaches that rely on the optimization solution, which is obtained with complete global information, may not be practically feasible and efficient.
	\item {\em Multiple Entities and Rationality:} IoT components may belong to or are operated by different entities, e.g., sensor owners, wireless service providers, and data center operators, and they have different objectives and constraints. System optimizations which support a single objective will fail to model and determine an optimal interaction among these self-interested and rational entities. 
	\item {\em Incentive Mechanism:} In addition to system performance and QoS requirement, from a business perspective, incentives such as cost, revenue, and profit are essential drivers to sustain the IoT development and operation. Therefore, the design and implementation of IoT services have to take incentive factors into account. This incentive issue becomes more complex when there are multiple entities interacting to achieve their own objectives. Incentive mechanisms have to be carefully designed to achieve not only maximal efficiency, but also stable and fair solutions among rational entities.
\end{itemize}

Therefore, economic approaches are considered as an alternative when designing and implementing IoT services. Economic approaches involve the analysis and optimization of the production, distribution, and consumption of goods and services. The approaches aim to analyze how IoT economies work and how IoT entities interact economically. In the following, we discuss important economic approaches and IoT related works.

\subsubsection{Cost-Benefit Analysis}

Cost-Benefit Analysis (CBA) is a method to estimate an equivalent money value in terms of benefits and costs from IoT systems and services. CBA involves computing the benefits against costs for the entities to make economic and technical decisions, for example, whether the system and service should be implemented or not, which technology and design should be adopted, and what the risk factors are. In~\cite{uckelmann2012}, the authors present the performance measurement and CBA for using RFID and IoT in logistic applications. In particular, the authors identify the cost and benefit of implementing RFID projects and justify the IoT investment for logistic company. CBA first determines the possible projects, designs, and their stakeholders. The metrics and cost/benefit elements are defined and calculated. Some important metrics considered are Total Cost of Ownership (TCO), Activity-Based Costing (ABC), Net Present Value (NPV), and Economic Value Added (EVA). Then, various costs are classified into different categories. For example, the physical world costs include the cost of RFID tags, the cost of applying the tags to products, and the cost of purchasing and deploying tag readers. The syntactics cost includes system integration cost, and the pragmatics cost includes the cost of implementing application solution. Next, the potential benefits are determined including the improved information sharing, reduced shrinkage, reduced material handling, and improved space utilization, etc. The stakeholders that receive the benefits are identified including manufacturers and suppliers, retailers, and consumers. Finally, the case study in the beverage supply chain is discussed, where actual money for costs and benefits are calculated and estimated. By using the CBA method, it is found that the benefits can be distributed among different parties, e.g., brewery (28.5\%), bottler (19.1\%), wholesaler (24.7\%), and retailer (27.6\%). Based on this observation, the authors introduce a simple Cost-Benefit Sharing (CBS) scheme that allows stakeholders to achieve different levels of benefits.

\subsubsection{User Utility}

From economics, utility represents the satisfaction and preference of consumers on choices of products or services. The concept of utility has been long and extensively used in computer networks and distributed computing to provide an abstraction of system performance perceived by users. For example, the satisfaction of network bandwidth is widely quantified by a concave utility function, e.g., the logarithmic function, which complies with the ``law of diminishing returns''. In particular, the rate of satisfaction increase decreases as the bandwidth becomes larger. Utility is adopted as an objective function for system optimizations meaningfully to maximize the users' satisfaction. In IoT, for example, utility is used to quantify the QoS performance of the sensor data collection system for smart city~\cite{alfagih2013}. The utility can be obtained from a survey data~\cite{jiang2005}. The system is composed of an access point that receives data from the stationary or mobile data collectors. The collectors gather sensing data from a number of sensors. The access point receives different types of data, e.g., delay-sensitive and delay-tolerant, with different QoS requirements. The utility for delay, sensing quality, and trust is defined based on exponential, sigmoid, and power functions, respectively. For example, when delay increases, the utility decreases exponentially. The access point then uses the information about utility to optimize the revenue of sensing data collection services.

Utility can be used further to determine good or service demand from users. Demand can be obtained as a function of price to indicate the amount of good or service consumed by the users that maximizes their utility. Let $U(q,p)$ denote the utility given that the users consume the good or service with amount $q$ and price $p$. The demand is obtained as $D(p) = \arg \max_q U(q,p) $. Based on this fact, service providers can set the price accordingly. 

\subsubsection{Market and Pricing} 

Markets are economic systems, procedures, social relations, and infrastructure established to support good and service exchange. The trade is made in the market that sellers offer goods or services to buyers who pay money to the sellers. Pricing is an essential mechanism of the market to ensure the efficiency of trading, i.e., sellers gain the highest profit while buyers maximize their satisfaction. 

IoT application markets are introduced in~\cite{munjin2012}. The authors in~\cite{munjin2012} highlight that the IoT application markets can imitate that of mobile application marketplace, e.g., Apple AppStore and Google Play. They also propose that the IoT application marketplace should focus at the data market, and introduce basic IoT marketplace structure. In the proposed marketplace, IoT devices are connected with a middleware and data broker. The data broker sells its data in the application markets of IoT marketplace. Buyers can purchase and use the data for their software applications. Nonetheless, the authors do not discuss the methods of pricing in the IoT marketplace.

In the literature, different approaches can be adopted for IoT service and data pricing.
\begin{itemize}
	\item {\em Market Equilibrium:} This approach considers demand from buyers and supply from sellers, respectively. The demand decreases while the supply increases as the price increases. The market equilibrium, which is similar to the Walrasian equilibrium in economics, is the point where supply equals demand. The authors in~\cite{chavali2012} adopt this market equilibrium pricing for IoT-based multi-modal sensor networks in a monopoly setting, i.e., one seller in the market. A sensor owner as a seller sells data to users which are buyers. The demand is determined as the users' preference of buying the sensor data that maximizes their utility given the budget. The supply is determined as the sensor owner's optimal strategy of selling the data that maximizes the profit given the cost of producing the data. The market is cleared and the equilibrium price is obtained when the demand and supply balance. The authors apply this pricing scheme to target tracking applications.
	\item {\em Duopoly and Oligopoly Market:} Duopoly and oligopoly are the market structures with two and more than two sellers, respectively. In duopoly and oligopoly markets, to maximize profits, sellers compete each other in terms of price or supply quantity, referred to as the Bertrand or Cournot competition models, respectively. Game theory is a useful tool for analyzing the Nash equilibrium solution. The authors in~\cite{Feng_2014_TC} study the monopoly and oligopoly markets of cloud resource pricing to support IoT services. Users choose a seller if their utility from using cloud resource minus the price is positive. If there are multiple sellers, the seller that yields the maximum utility minus price is selected by the users. The authors study important properties of the Nash equilibrium prices, e.g., the existence of the Nash equilibrium.
	\item {\em Auction:} Auction can be used as a pricing mechanism for IoT services. There are different types of auctions, e.g., single and double-side auction. In the single-side auction, one seller auctions good or service by requesting bids from multiple buyers, or one buyer receives asks from multiple sellers and chooses the best seller. Alternatively, in the double-side auction, multiple sellers and buyers submit their asks and bids, and the auctioneer determines sets of winner sellers and buyers, clearing price, and good or service allocation. More details of auction and its applications in data communications can be found from~\cite{zhang2013}. Auctions are also adopted in IoT services~\cite{cao2015} particularly for crowdsourcing of target tracking applications. The fusion center needs to collect sensing data from different sensors to determine a state of the target. Thus, the fusion center requests for bids from sensors and chooses the winning sensors to buy the sensing data from. The solution of the auction is obtained from solving the multiple-choice knapsack problem to achieve maximum utility for the fusion center.
\end{itemize}

\begin{figure}
\begin{center}
$\begin{array}{c} \epsfxsize=4.6 in \epsffile{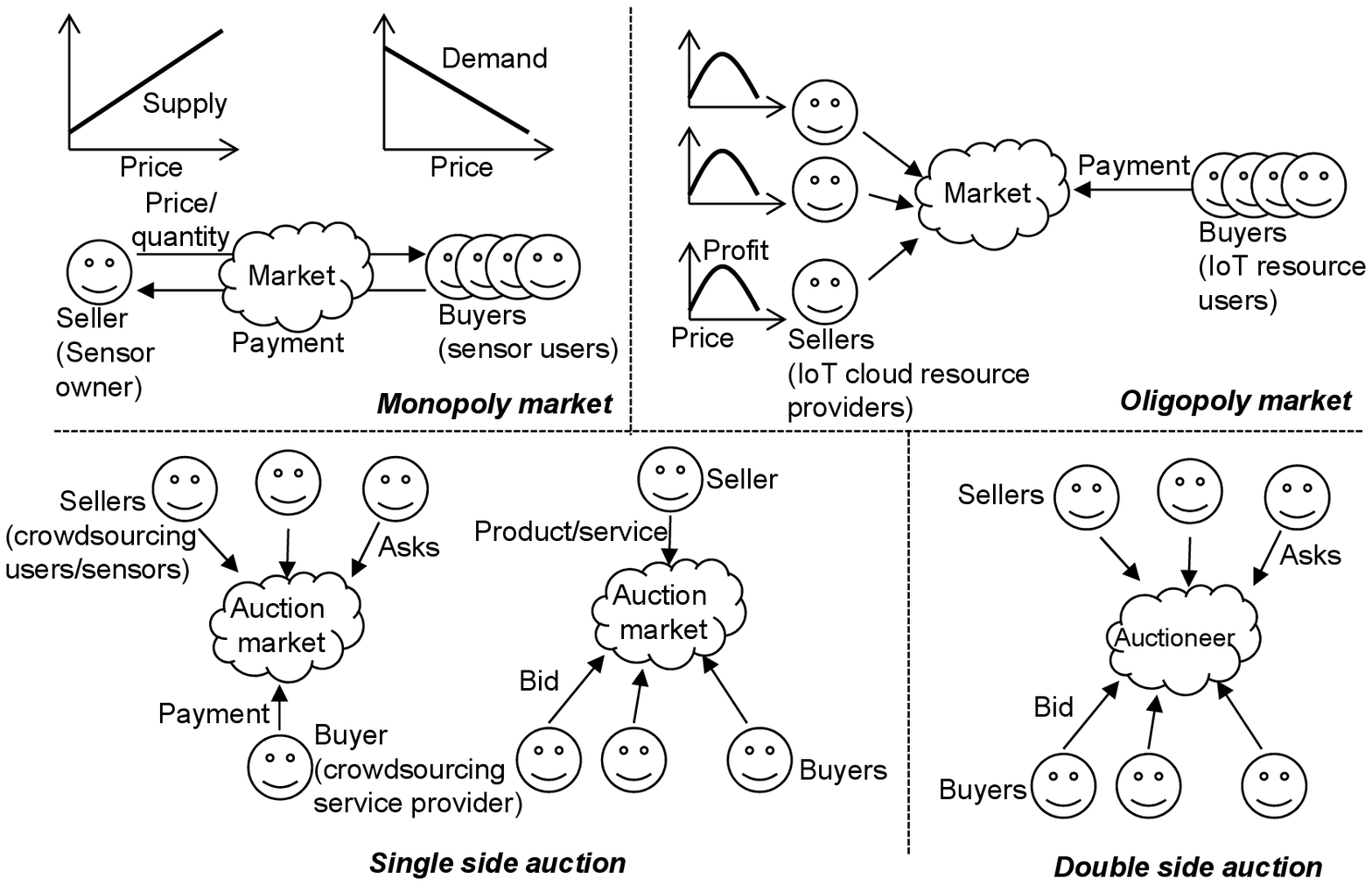} \\ [-0.2cm]
\end{array}$
\caption{Different market structures.} 
\label{fig:market}
\end{center}
\end{figure}

\vspace{0.2in}
\noindent{\bf Callout: Figure~\ref{fig:market} Different market structures.}
\vspace{0.2in}

Figure~\ref{fig:market} shows the different market structures applicable to IoT. In addition to sensor data and cloud services, there are some other resources and services in IoT that can traded by adopting market and pricing mechanisms. 
\begin{itemize}
	\item {\em Energy} is used to power a variety of IoT components, e.g., sensors, data gateways, base stations, data centers, backbone and edge networks. In smart grid, energy can be traded in utility markets~\cite{erolkantarci2015}. In monopoly or oligopoly markets, utility company(ies) can optimize the prices of energy supplied to data centers, wired and wireless networks to maximize their profits given energy demands. 
	\item {\em Spectrum and network bandwidth} are scarce resources, especially in wireless networks. In cognitive radio networks, spectrum can be traded in a market in a highly dynamic fashion. Specifically, licensed users can sell their free spectrum to unlicensed users to earn more revenue and improve spectrum utilization. Various trading models have been introduced including auctions~\cite{zhang2013}. 
	\item {\em Data and information services} can be offered and integrated to support IoT applications. Such services are, for example, information searching, data storage and mining, and information security protection. The concept of ``anything as a service (XaaS)'' is recently introduced that allows any resources to be treated and used as services. The typical ones are Software as a Service (SaaS) and Monitoring as a Service (MaaS). The authors in~\cite{xu20xx} introduce using a contract theory to study data mining services that allow data owners to sell their data to the data collector. To protect privacy of data owners, the data collector performs data anonymization, and resells the anonymized data to data miners. The data collector optimizes choices of contracts based on data quality, privacy requirement, and payment proposed to the data owners so that the profit is maximized.
\end{itemize}
In IoT, data and information can be treated as resources and services that have to be optimized, especially to maximize their utilization, revenue, and profit of owners and providers. In the next section, we will present an overview of information economics that can be applied to IoT. Note that we use ``data'' and ``information'' interchangeably in the rest of the paper to simplify explanation despite their subtle difference.

\section{Introduction to Information Economics}
\label{sec:infoecon}

Information economics focuses on various aspects of information in economy. Information has a unique feature that it can be easily created, but possesses diverse levels of reliability and trust. Information can be used to make a decision in various problems. Thus, the value of information has to be determined. In this section, we briefly discuss two major aspects of information economics, i.e., the value of information and information good pricing.

\subsection{Value of Information}

Information is used in decision making to achieve the goal of systems and services. Information can change the knowledge of a decision maker on a particular subject. Let the knowledge be represented by a probability distribution of state $x$. If information $y$ is used in decision making that yields payoff to the decision maker. The value of information is defined as follows~\cite{lawrence1999}: $v(x,y) = \pi (x, a_y) - \pi ( x, a_0)$, where $\pi(x,a)$ is the payoff given state $x$ and decision $a$. $a_y$ and $a_0$ are the decisions after and before having information $y$, respectively. The value of information can be positive, zero, or negative, depending on the quality of the information. 

Value of information facilitates system design in the following aspects~\cite{lawrence1999}.
\begin{itemize}
	\item {\em Optimal Decision:} Given the knowledge of system states, an optimal decision can be made to maximize expected payoff which is defined as follows: $\max_a \int_{\mathbb{X}} \pi(x,a) p(x | y) {\mathrm{d}} x$, where ${\mathbb{X}}$ is the state space, $p(x|y)$ is the probability distribution of state $x$ conditioned on the available information $y$. 
	\item {\em Information Source Selection:} Since the payoff depends on the information $y$, its source has to be evaluated and optimized. In IoT, there can be many sensors performing similar sensing tasks. The information from the sensor that yields the highest value of information, i.e., making the best decision should be chosen.
	\item {\em Information System Optimization:} However, collecting information to make an optimal decision also incurs a certain cost. In IoT, the sensors consume energy and bandwidth to collect and transfer sensing information. Information processing uses computing resource from cloud services. Therefore, the information system optimization is important to measure all the costs and trade off with the value of information. The difference between value and cost is called information gain that should be maximized for the designed information system.
\end{itemize}

Value of information analysis has been applied to sensor networks, which are an essential part of IoT. For example, the authors in~\cite{turgut2013} analyze the value of information in energy-constrained intruder tracking sensor networks. The value of information depends on the damage that can be avoided per having additional information. The damage is defined as a function of tracking error. The value of information is the difference between the maximum damage when the tracking sensor network is not deployed, and the actual damage if tracking information is available. The same authors show the usefulness of the analysis to optimize data transmission for underwater sensor nodes such that the value of information is maximized. In particular, an autonomous underwater vehicle (AUV) travels to collect data from sensors. Not only the decision on which and when sensors should transmit data about an intruder, but also the traveling path of the AUV to collect sensor data are optimized.

\subsection{Information Good Pricing}

Another aspect of information economics is treating information as intangible good to be sold in a market. Information good has unique characteristics.
\begin{itemize}
	\item {\em Quality Dependent Good:} Consumers value information differently. The value of information depends on the reputation and quality more than quantity. Additionally, as aforementioned, the value could depend on the improvement of decision making and consequently sources of information. 
	\item {\em Different Cost Structure:} There are different levels of costs. The fixed cost incurred from information system design, development, and deployment is higher than the variable cost for producing information, and they are higher than the cost for reproducing and storage. For example, deploying sensors and communication infrastructure requires hefty investment. 
	\item {\em Versioning and Bundling:} Information can be offered in different version, e.g., with different quality. For example, IoT sensing data can be offered with different resolution depending on application requirements. Additionally, different sets of information can be bundled to enhance their values. In IoT, multi-modal sensor data, e.g., from motion detection and video camera for surveillance applications, can be used jointly to improve the detection performance. 
\end{itemize}

Because of the unique features of the information good, pricing mechanisms for selling information have to be developed differently from those of other tangible goods. In IoT context, the following issues can be studied.

\begin{itemize}
	\item {\em Choices of Pricing:} To obtain IoT sensing information and services, different choices of pricing can be employed. Transaction based pricing charges users when they access information and services. Information/time unit pricing charges users according to the amount of information or time taken to access services. Subscription based pricing allows users to access information and services for a certain time period. For example, transaction based pricing could be suitable for sensing information search, while information/time unit pricing is suitable for streaming sensing data, e.g., video.
	\item {\em Profit and Cost Optimization:} As typical for the IoT information and service provider, the profit, i.e., revenue minus cost, must be maximized. This can be achieved through different approaches. As investment accounts for major cost of IoT, IoT infrastructure has to be optimally designed and deployed, e.g., how many and where sensors, gateways, base stations, and data centers should be deployed. Sensing information collection and service delivery incur a certain cost, e.g., energy and network bandwidth. The quality of information and service, which affects resource usage and demand, can be optimized jointly with price to achieve the maximum profit.
	\item {\em Price Competition:} It is common to have multiple IoT information and service providers in the market, and thus competition is inevitable. Game theory is used to analyze pricing strategies of service providers. However, new game models have to be developed taking unique characteristics of information into account. For example, they have to incorporate different choices of pricing and different cost structures. Moreover, competition is affected by information demand, which depends on the value perceived by users. 
\end{itemize}
In the next section, we propose a simple game theoretic model to analyze the price competition for IoT sensing information. 


\section{IoT Service Competition}
\label{sec:competition}

In this section, we demonstrate an example model of information economics to address the IoT service pricing competition. We first describe the system model of the IoT sensing information market. Then, we present a simple noncooperative game formulation. Some numerical results and outlooks for possible extensions are discussed afterward.

\subsection{Sensing Information Market}

We consider $S$ IoT services which competitively sell event detection or environmental sensing information to users. Without loss of generality, the sensing information is binary, i.e., indicating whether an event happens or not. However, a sensing error can occur. The detection probability of service $s$ is denoted by $P_{\mathrm{d}}(s)$, and thus the miss detection probability is $1-P_{\mathrm{d}}(s)$. The false alarm probability is denoted by $P_{\mathrm{f}}(s)$. The miss detection is an error that an event happens, but the sensing information reports no event. By contrast, the false alarm is an error that an event does not happen, but sensing information indicates the presence of the event. 

Users can buy sensing information to be used for their own applications. All the users are charged by the price $p(s)$ for buying sensing information from service $s$. The users can buy sensing information from a single service or multiple services. For the former, the user regards the sensing information as the ``substitute good'' that the user can switch to buy from the best service. On contrary, for the latter, the users treat the sensing information as the ``complementary good'' that if the user needs, it has to buy from all services. When the users buy sensing information from multiple services, the information can be combined, i.e., fusion, to obtain better sensing accuracy. Accordingly, the users will pay all the services that they buy the sensing information from. We consider two common fusion rules, i.e., OR and AND. For the OR fusion rule, if one of the sensing information indicates that there is an event, the user will conclude that the event happens. By contrast, for the AND fusion rule, all of the sensing information must indicate that there is an event, so that the user will conclude that the event happens.

\begin{figure}
\begin{center}
$\begin{array}{c} \epsfxsize=4.5 in \epsffile{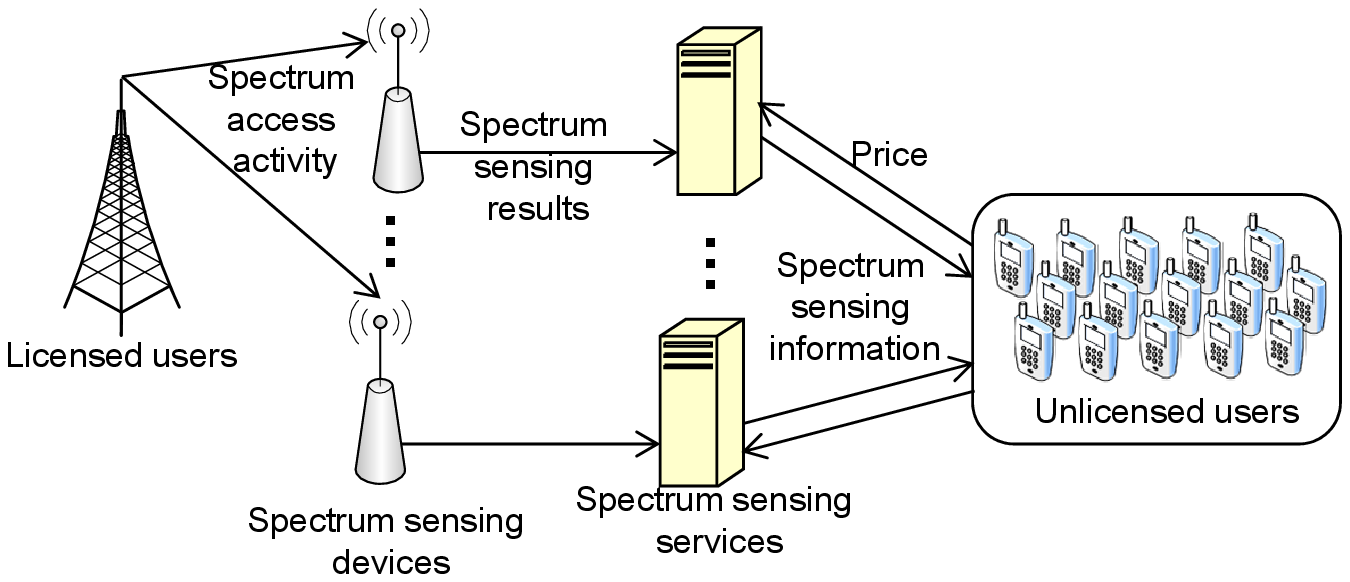} \\ [-0.2cm]
\end{array}$
\caption{Spectrum sensing service example.} 
\label{fig:spectrumsensing}
\end{center}
\end{figure}

\vspace{0.2in}
\noindent{\bf Callout: Figure~\ref{fig:spectrumsensing} Spectrum sensing service example.}
\vspace{0.2in}

One example of the sensing information market is in cognitive radio networks. Spectrum sensing networks composed of spectrum sensor devices can be deployed by third parties as IoT services to monitor spectrum activity on a certain band. The spectrum sensing services can sell their spectrum availability information to unlicensed users for dynamic spectrum access (Fig.~\ref{fig:spectrumsensing}). The unlicensed users can choose to buy spectrum availability information from different sensing networks which charge different prices.

\subsection{Game Formulation}

Given the above IoT (sensing) services, we wish to study the competition in setting the price of sensing information. For the substitute case, the user buys sensing information from one service. The utility of the user buying from service $s$ is defined as follows:
\begin{equation}
	U(s) = v P_{\mathrm{d}}(s) - P_{\mathrm{f}}(s) - p(s)	,
\end{equation}
where $v$ is the weight of the detection probability relative to the false alarm probability and price. The weight is random in the user population following a certain distribution, e.g., uniform. The demand of sensing information from service $s$ is generated by a user when the utility is the highest and above zero. It is denoted by $D_s({\mathbf{p}})$ where ${\mathbf{p}} = (p_1,\ldots,p_S)$ contains the prices of all $S$ services.

For the complementary case, the user buys sensing information from all services, the set of which is denoted by ${\mathcal{S}}$. The utility of the user is defined as follows:
\begin{equation}
	U({\mathcal{S}}) = v P_{\mathrm{d}}({\mathcal{S}}) - P_{\mathrm{f}}({\mathcal{S}}) - \sum_{s \in {\mathcal{S}} } p(s)	,
\label{eq:utility_fusion}
\end{equation}
where $P_{\mathrm{d}}({\mathcal{S}})$ and $P_{\mathrm{f}}({\mathcal{S}})$ are the detection and false alarm probabilities from a certain fusion rule, respectively. Again, the demand is generated when the utility is higher than zero.

Now, we present the noncooperative game formulation of price competition among IoT services selling sensing information to users. The players of the game are the services. Their strategies are the prices. The payoff is the profit which is defined as $F_s ( p(s), {\mathbf{p}}_{-s} ) = p(s) D_s( {\mathbf{p}} ) - C_s$, where $C_s$ is the cost of generating sensing information of the service $s$. Because of the unique nature of the information which can be reproduced and transferred to users without a cost, this cost is a constant and is independent of demand and price strategy. ${\mathbf{p}}_{-s}$ contains the prices of all services except service $s$. Then, the best response of the player is the price that yields the highest payoff, i.e., $p_s^*( {\mathbf{p}}_{-s} ) = \max_{p(s)} F_s ( p(s), {\mathbf{p}}_{-s} )$ and the Nash equilibrium is $p^*_s( {\mathbf{p}}^*_{-s} )$ for all services. Here, the Nash equilibrium is the set of prices that none of services can deviate unilaterally to gain a higher profit.

\subsection{Numerical Results}

We show the numerical results to demonstrate the sensing information pricing. To ease the presentation of results, we consider two services selling sensing information to a population of users. We consider two cases, i.e., substitute and complementary services, that the users buy sensing information from one of services or from both of services, respectively. The weight of detection probability is uniformly distributed within the range $[0,2]$. The detection probabilities of services 1 and 2 are 0.8 and 0.9, and the false alarm probabilities are 0.1 and 0.2, respectively.

\subsubsection{Substitute Services}

\begin{figure}
\begin{center}
$\begin{array}{c} \epsfxsize=4.0 in \epsffile{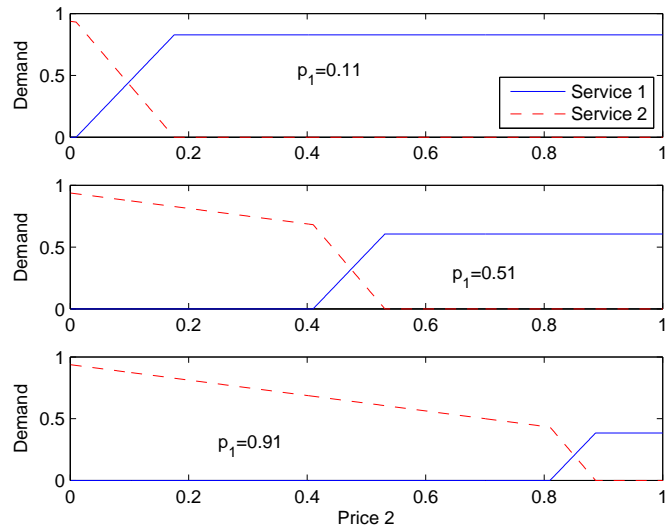} \\ [-0.2cm]
\end{array}$
\caption{Demand of two substitute services.} 
\label{fig:demand_single}
\end{center}
\end{figure}

\vspace{0.2in}
\noindent{\bf Callout: Figure~\ref{fig:demand_single} Demand of two substitute services.}
\vspace{0.2in}

Figure~\ref{fig:demand_single} shows the demand for substitute services 1 and 2 when their prices are varied. In particular, we consider three prices of service 1, i.e., 0.11, 0.51, and 0.91, which correspond to the cases of low, medium, and high prices, respectively. We make the following observation on the demand. Firstly, with substitute services, when the price of one service increases, the utility of the users buying that service will decrease. Consequently, the user will compare the utility received from alternative services and switch to the one that yields the highest utility. Thus, the demand of the service with the increased price will decrease, while the demand of the other service will increase. Secondly, we observe that there are three parts of the demand of service 2. In the first part, the demand decreases slowly. This corresponds to the case that the price of service 2 is high such that some users have negative utility, and thus they will deviate from buying information from any service. In the second part, the demand decreases sharply. This corresponds to the case that some users find that choosing to buy information from service 1 yields a higher utility. Thus, the demand of service 1 also increases. In the third part, the price of service 2 is too high that all users will choose to buy information from service 1. Thus, the demand of service 2 is zero. 

\begin{figure}
\begin{center}
$\begin{array}{c} \epsfxsize=4.0 in \epsffile{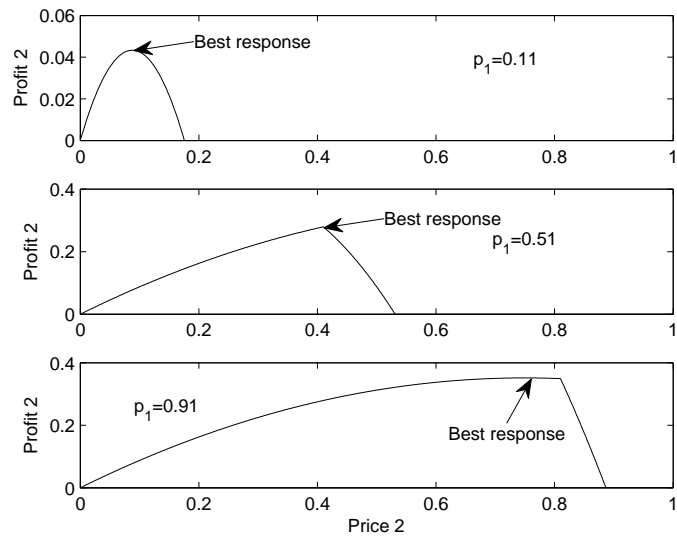} \\ [-0.2cm]
\end{array}$
\caption{Profit of service 2 under two substitute services.} 
\label{fig:profit_single}
\end{center}
\end{figure}

\vspace{0.2in}
\noindent{\bf Callout: Figure~\ref{fig:profit_single} Profit of service 2 under two substitute services.}
\vspace{0.2in}

\begin{figure}
\begin{center}
$\begin{array}{c} \epsfxsize=4.0 in \epsffile{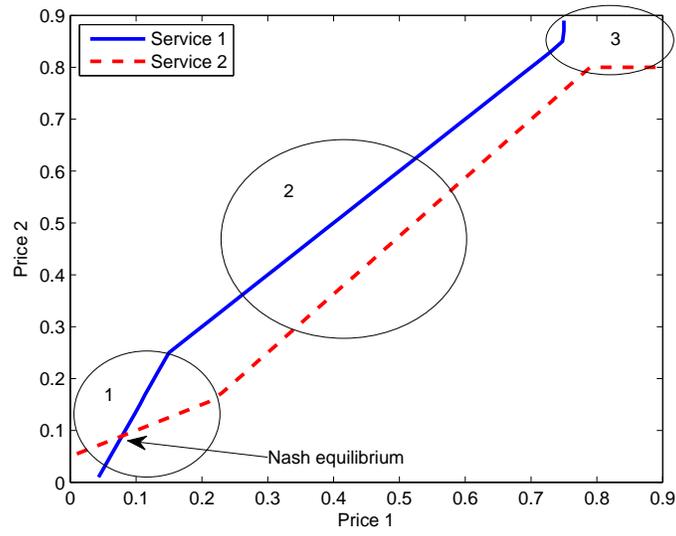} \\ [-0.2cm]
\end{array}$
\caption{Best responses and Nash equilibrium of two substitute services.} 
\label{fig:nash_single}
\end{center}
\end{figure}

The profit of service 2 increases first and then decreases as the price of service 2 increases as shown in Fig.~\ref{fig:profit_single}. The highest profit is the best response in terms of price. We observe that different structures of demand result in different profit of service 2. This is clearly shown in Fig.~\ref{fig:nash_single}, which illustrates the best responses of two services. In the first part, the best response depends on the price that yields the demand to maximize the profit of service 2. In the second part, the best response depends on the price that users start to switch to service 1. In the third part, the best response is not affected by the price of service 2 as it is too high, and thus the best response remains constant.

\vspace{0.2in}
\noindent{\bf Callout: Figure~\ref{fig:nash_single} Best responses and Nash equilibrium of two substitute services.}
\vspace{0.2in}

From Fig.~\ref{fig:nash_single}, the intersection between the best responses of services 1 and 2 is the Nash equilibrium prices. It is possible to show that the Nash equilibrium for information selling among substitute services always exists and is unique. The proof similar to that in~\cite{sarvary1997} can be applied.

\subsubsection{Complementary Services}

\begin{figure}
\begin{center}
$\begin{array}{c} \epsfxsize=4.0 in \epsffile{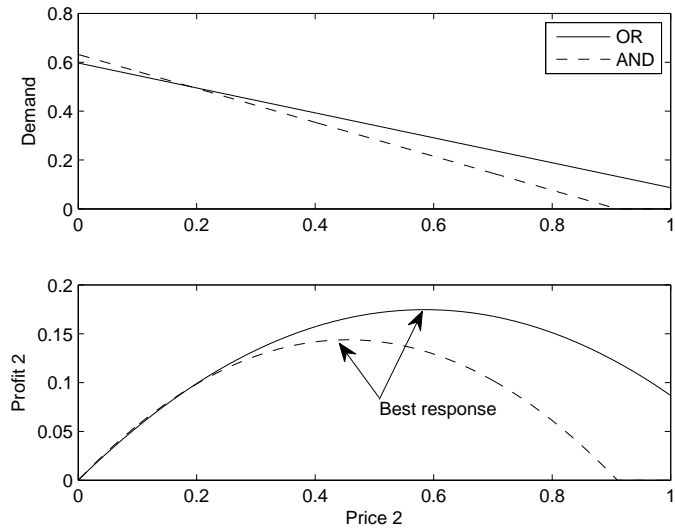} \\ [-0.2cm]
\end{array}$
\caption{Demand and profit of service 2 under two complementary services.} 
\label{fig:demand_fusion}
\end{center}
\end{figure}

\vspace{0.2in}
\noindent{\bf Callout: Figure~\ref{fig:demand_fusion} Demand and profit of service 2 under two complementary services.} 
\vspace{0.2in}

For comparison, we consider the complementary services. Figure~\ref{fig:demand_fusion} shows the demand and profit when the OR and AND fusion rules are adopted. Here note that since the users are indifferent about the prices from any complementary services as given in (\ref{eq:utility_fusion}), the demands of all services are equal. Unlike substitute services, when the price of one service decreases, the demand of both services decreases, since the users want to buy information from all the services. If one of them is expensive, the users will buy less from both. Additionally, we observe that the demand of the OR fusion rule decreases more slowly than the AND fusion rule, and also the best response of the AND fusion rule is smaller. This is due to the fact that with the OR fusion rule, the improvement from higher detection probability is more significant than the degradation from the higher false alarm probability.

\begin{figure}
\begin{center}
$\begin{array}{c} \epsfxsize=4.0 in \epsffile{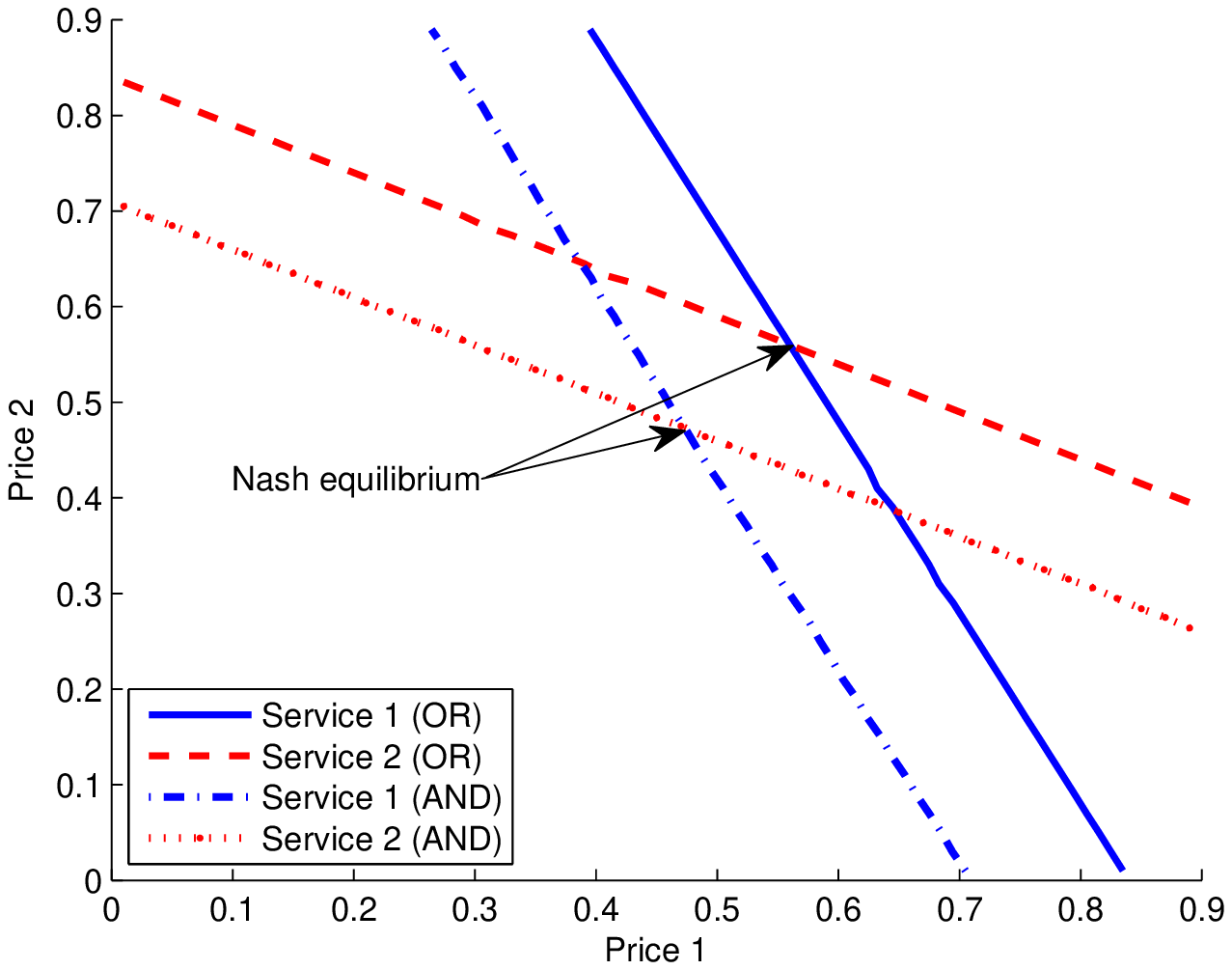} \\ [-0.2cm]
\end{array}$
\caption{Best responses and Nash equilibrium of two complementary services.} 
\label{fig:nash_fusion}
\end{center}
\end{figure}

\vspace{0.2in}
\noindent{\bf Callout: Figure~\ref{fig:nash_fusion} Best responses and Nash equilibrium of two complementary services.} 
\vspace{0.2in}

Figure~\ref{fig:nash_fusion} shows the best responses of both services when the OR and AND fusion rules are applied. The Nash equilibrium prices of the AND fusion rule are lower than those of the OR fusion rule. Moreover, the Nash equilibrium prices of the complementary services are higher than those of the substitute services, and thus achieving the higher profit. This is intuitive because with substitutability, the competition between the services is severe and they have to lower the price to gain the highest profit. By contrast, with complementarity, the competition is mild and they do not have to reduce the price too much.

\subsection{Extensions}



Placemeter (http://www.placemeter.com/) offers to buy sensing data, i.e., video from a camera, capturing picture from streets in New York. It is advertised that the users can be paid up to \$50 a month depending on the quality of video, e.g., picture angle. Placemeter performs video analytics on the data to provide useful and sellable information, for example, to help business identify groups of potential consumers and to make tourists aware of waiting lines at shops and museum. In the market model of Placemeter, multiple users, i.e., sellers, set up their cameras, capture video in a similar area, and sell the video data to Placemeter, i.e., a buyer. In this case, Placemeter has choices of acquiring the video data from different sources, which may provide similar view from the city. Thus, the sellers are facing a competitive situation to set the price to be attractive for the buyer. Depending on the quality of the video data, the buyer will quantify the utility and derive the information demand that will affect the competitive pricing of the sellers. Our proposed information economic framework will be useful in analyzing this situation.

In this example, clearly the market structure and pricing mechanism are different from computer networks. The possible arising issues are as follows:
\begin{itemize}
	\item {\em Value of information} has to be quantified. Different video data, after being processed, can yield different useful information. The utility of the video from different users has to be measured, and afterward the demand can be derived. Unlike in computer networks, the new utility and demand functions have to be proposed to be suitable for information goods. Substitutability and complementarity of the video data will be incorporated. Video pictures from a similar angle can be substitute as they yield similar information extraction performance. By contrast, video pictures from different angles can be complementary as they can be used jointly to improve the information extraction effectiveness.
	\item {\em Information re-selling} is possible. In this case, Placemeter can process video data and sell it to other businesses or customers. The difference between the price paid to the users selling the video data and the price obtained from the businesses or customers will be the revenue for Placemeter. Thus, it is important to quantify the utility of the video data so that the revenue is maximized.
	\item {\em Competition} will arise when there are multiple information buyers, i.e., competitors to Placemeter. In this situation, the users have choices to sell information to multiple buyers. Thus, it is important to set prices accordingly. For example, the prices of video data can be lower when there are multiple buyers as the sellers can gain more revenue from multiple buyers without or with a marginal cost of additional information capture and transfer. This hypothesis can contradict with the well-known result. In traditional markets, similar to that in computer networks, when there are multiple buyers, the prices should increase because of higher demand. Therefore, a new game theoretic model needs to be developed to analyze such information selling competition.

\end{itemize}

\subsection{Future Work}

Based on the proposed sensing service pricing model, the following extensions can be pursued.
\begin{itemize}
	\item {\em Impact of Sensing Information Correlation:} Sensing information from different services can be correlated. Users can selectively buy sensing information from only some of available services to lower their cost. The model to analyze the impact of the correlation and pricing can be built.
	\item {\em Collusion and Price of Anarchy:} Multiple services can collude to optimize their prices and obtain the highest profit. The collusion among services and the price of anarchy when the services adopt optimal pricing scheme can be investigated.
	\item {\em Time Sensitive Information:} Information can be time sensitive. Its demand depends on time, e.g., utility of sensing information decreases as a function of time after when it is generated. A dynamic game model, which takes a time parameter into account, is a candidate to analyze this situation.
	\item {\em Information Resell:} Information can be reproduced virtually without cost. Some users may buy information, copy, and resell to other users. This introduces a hierarchy in the information market. Hierarchical games, e.g., Stackelberg games, can be applied to this case.
\end{itemize}


\section{Conclusion}
\label{sec:conclusion}

Internet of Things has emerged as a promising technology to connect devices and provide services. In this article, we have considered economics of IoT that is an important aspect beyond system optimizations. We have first presented different economic approaches to address a variety of issues in IoT. We have then particularly considered value of information and information good pricing directions. To demonstrate the application of information economics, we have presented the game theoretic model for sensing information price competition. The model considers both the substitute and complementary services. The solution in terms of the Nash equilibrium has been obtained. Finally, important research directions are outlined.

\section*{Acknowledgements}

This work was supported in part by the National Research Foundation of Korea (NRF) grant funded by the Korean government (MSIP) (2014R1A5A1011478), Singapore MOE Tier 1 (RG18/13 and RG33/12) and MOE Tier 2 (MOE2014-T2-2-015 ARC 4/15), and the U.S. National Science Foundation under Grants US NSF CCF-1456921, CNS-1443917, ECCS-1405121, and NSFC 61428101.

\newpage

\begin{thebibliography}{99}


\bibitem{atzori2010}
L. Atzori, A. Iera, and G. Morabito, ``The Internet of Things: A survey,'' {\em Computer Networks}, vol. 54, no. 15, pp. 2787-2805, October 2010.

\bibitem{gluhak2011}
A. Gluhak, S. Krco, M. Nati, D. Pfisterer, N. Mitton, and T. Razafindralambo, ``A survey on facilities for experimental internet of things research," {\em IEEE Communications Magazine}, vol. 49, no. 11, pp. 58-67, November 2011.


\bibitem{xu2014}
J. Xu, Y. Andrepoulos, Y. Xiao, and M. van der Schaar, ``Non-stationary resource allocation policies for delay-constrained video streaming: Application to video over Internet-of-Things-enabled networks,'' {\em IEEE Journal on Selected Areas in Communications}, vol. 32, no. 4, pp. 782-794, April 2014.

\bibitem{uckelmann2012}
D. Uckelmann, ``Performance measurement and cost benefit analysis for RFID and Internet of Things implementations in logistics,'' {\em Quantifying the Value of RFID and the EPCglobal Architecture Framework in Logistics}, pp. 71-100, 2012.

\bibitem{alfagih2013}
A. E. Al-Fagih, F. M. Al-Turjman, W. M. Alsalih, and H. S. Hassanein, ``A priced public sensing framework for heterogeneous IoT architectures,'' {\em IEEE Transactions on Emerging Topics in Computing}, vol. 1, no. 1, pp. 133-147, June 2013.

\bibitem{jiang2005}
Z. Jiang, Y. Ge, and Y. Li, ``Max-utility wireless resource management for best-effort traffic,'' {\em IEEE Transactions on Wireless Communications}, vol. 4, no. 1, pp. 100-111, January 2005.


\bibitem{munjin2012}
D. Munjin and J. Morin, ``Toward Internet of Things application markets," {\em IEEE International Conference on Green Computing and Communications (GreenCom)}, pp. 156-162, November 2012.



\bibitem{chavali2012}
P. Chavali and A. Nehorai, ``Managing multi-modal sensor networks using price theory,'' {\em IEEE Transactions on Signal Processing}, vol. 60, no. 9, pp. 4874-4887, September 2012.

\bibitem{Feng_2014_TC}
Y. Feng, B. Li, and B. Li, ``Price competition in an oligopoly market with multiple IaaS cloud providers,'' \emph{Transactions on Computers}, vol. 63, no. 1, Jan. 2014.

\bibitem{zhang2013}
Y. Zhang, C. Lee, D. Niyato, and P. Wang, ``Auction approaches for resource allocation in wireless systems: A survey,'' {\em IEEE Communications Surveys and Tutorials}, vol. 15, no. 3, pp. 1020-1041, Third Quarter 2013.

\bibitem{xu20xx}
L. Xu, C. Jiang, Y. Chen, Y. Ren, and K. J. R. Liu, ``Privacy or utility? A contract theoretic approach to data collection,'' {\em IEEE Journal of Selected Topics in Signal Processing}, to appear.

\bibitem{cao2015}
N. Cao, S. Brahma, and P. K. Varshney, ``Target tracking via crowdsourcing: A mechanism design approach,'' {\em IEEE Transactions on Signal Processing}, vol. 63, no. 6, pp. 1464-1476, March 2015.

\bibitem{erolkantarci2015}
M. Erol-Kantarci and H. T. Mouftah, ``Energy-efficient information and communication infrastructures in the smart grid: A survey on interactions and open issues,'' {\em IEEE Communications Surveys \& Tutorials}, vol. 17, no. 1, pp. 179-197, Firstquarter 2015.




\bibitem{lawrence1999}
D. B. Lawrence, {\em The Economic Value of Information}, Springer, April 1999.

\bibitem{turgut2013}
D. Turgut and L. Boloni, ``IVE: Improving the value of information in energy-constrained intruder tracking sensor networks,'' {\em IEEE International Conference on Communications (ICC)}, pp. 6360-6364, June 2013.



\bibitem{sarvary1997}
M. Sarvary and P. M. Parker, ``Marketing information: A competitive analysis,'' {\em Marketing Science}, vol. 16, no. 1, pp. 24-38, 1997.











\end{thebibliography}
\end{document}